**Experimental discovery of structure-property relationships in ferroelectric materials via active learning**

Yongtao Liu,[1] Kyle P. Kelley,[1] Rama K. Vasudevan,[1] Hiroshi Funakubo,[2] Maxim A. Ziatdinov,[1,3,a] and Sergei V. Kalinin[1,b]

[1] Center for Nanophase Materials Sciences, Oak Ridge National Laboratory, Oak Ridge, TN 37831

[2] Department of Material Science and Engineering, Tokyo Institute of Technology, Yokohama 226-8502, Japan

[3] Computational Sciences and Engineering Division, Oak Ridge National Laboratory, Oak Ridge, Tennessee 37831, USA

Emergent functionalities of structural and topological defects in ferroelectric materials underpin an extremely broad spectrum of applications ranging from domain wall electronics to high dielectric and electromechanical responses. Many of these have been discovered and quantified via local scanning probe microscopy methods. However, the search for these functionalities has until now been based by either trial and error or using auxiliary information such as topography or domain wall structure to identify potential objects of interest based on the intuition of operator or preexisting hypotheses, with subsequent manual exploration. Here, we report the development and implementation of a machine learning framework that actively discovers relationships between local domain structure and polarization switching characteristics in ferroelectric materials encoded in the hysteresis loop. The hysteresis loops per se and their scalar descriptors such as nucleation bias, coercive bias, hysteresis loop area, or more complex functionals of hysteresis loop shape and corresponding uncertainties are used to guide the discovery via automated piezoresponse force microscopy (PFM) and spectroscopy experiments. As such, this approach combines the power of machine learning methods to learn the correlative relationships between high dimensional data, and human-based physics insights encoded in the acquisition function. For ferroelectric, this automated workflow demonstrates that the discovery path and sampling points of on-field and off-field hysteresis loops are largely different, indicating the on-field and off-field hysteresis loops are dominated by different mechanisms. The proposed approach is universal and can be applied to a broad range of modern imaging and spectroscopy methods ranging from other scanning probe microscopy modalities to electron microscopy and chemical imaging.

[a] ziatdinovma@ornl.gov
[b] sergei2@ornl.gov

The rapid evolution of scanning probe and electron microscopy techniques over the last three decades has revolutionized the areas of science ranging from materials and condensed matter physics to chemistry and biochemistry.[1-5] As such, the various microscopy imaging modes have now become a mainstay across virtually all scientific fields. Similarly, the combination of imaging and spectroscopic modes in these techniques has provided a wealth of information on structure-property relations in these dissimilar systems. Examples include scanning tunneling microscopy and spectroscopy,[6-8] dark and bright field imaging in electron microscopy and electron energy loss spectroscopies,[9-11] topographic imaging and force-distance curve measurements in atomic force microscopy,[12-14] and electromechanical hysteresis loop measurements in piezoresponse force microscopy.[15-17] These structure-property relationships in turn yield a wealth of information on the underpinning physical, chemical, and biological mechanisms.

Very often the locations for spectroscopic measurements are selected manually based on the perceived (by human operator) interest of specific locations, as identified via features in a structural image. This point-and-click selection can be based on field-specific intuition, curiosity, and in special cases on a specific hypothesis. Alternatively, the measurements can be performed in the spectroscopic grid modes, where the spectral data is collected over a uniform sampling grid.[6,18-20] These in turn necessitated development of the linear and non-linear dimensionality reduction methods for analysis of such multidimensional data,[20,21] ushering exploratory machine learning methods into imaging areas. However, these imaging modalities are characterized by significant disparities in acquisition times for the spectroscopic and structural measurements. Correspondingly, the spatial density of the information is limited. While post-acquisition pan-sharpening methods based on compressed sensing, Gaussian process, etc. have been developed,[22,23] these approaches do not change the fundamental limitation of the spectroscopic imaging methods. Similarly, correlative learning of structure property relationships implemented via *im2spec* approach requires the availability of the full data set,[24] and implicitly assumes that the material properties did not change as the result of measurements.

The rapid progress in the computer vision methods enabled by the advent of the deep learning a decade ago[25] as well as wave of interest towards autonomous driving systems have stimulated strong interest in autonomous microscopy, with several notable opinion pieces over the last 3 years.[26-28] However, realization of this vision necessitates solution of three intertwined problems, including direct control of the microscope operation via external electronics, development of machine learning algorithms enabling the automated experiment (AE), and, perhaps less obviously, identifying the specific problems that AE seeks to resolve. Until now, this last problem has been largely overshadowed by the first two.

The direct control of microscopes has been available for decades, typically developed in the context of atomic and particle manipulation.[29-33] For imaging, the adaptive non-rectangular scanning approach was demonstrated by Ovchinnikov et al in 2009.[34] More recently, Huang et al[35] and Stores et al[36] have demonstrated the combination of machine learning algorithm with Atomic Force Microscopy (AFM) enabling autonomous operation of AFM without the need of human intervention in imaging modes, with the AE playing the role of (pretrained) feature identifier.

However, the second key component of the AE are the strongly coupled problem of machine learning algorithms and specific physical problem. Generally, the AE targeting the mechanisms of the ferroelectric domain wall pinning on structural defects will pursue a different strategy the experiment exploring the interaction of the ferroelectric and ferroelastic domain walls. Recently, we have reported the detailed analysis of the ML perspectives in automated/autonomous experiments in microscopy.[37] In particular, we noted that the AE itself is defined only in the context of prior knowledge, and seeks to discover new information or minimize uncertainties in the known system behavior. This coupled machine learning and physics problem in the context of active learning makes the AE a highly domain specific problem.

To date, the AE's have been implemented either using the human-based features engineering, or simple DCNN based image recognition.[36] For example, in piezoresponse force microscopy (PFM), the AE was introduced based on a line-by-line feedback system employed during classical rectangular scanning by Kelley *et al*.[38], termed "FerroBot". The FerroBot has shown the feasibility of the AE in PFM by using simple operator-defined features of interest. Finally, a Gaussian Process/Bayesian Optimization framework has been recently developed to control the probe trajectory via leveraging the explored information and scanning sequence.[39] At the same time, the classical Gaussian Process/Bayesian Optimization routines that underpin most AEs over the past years are based purely on the data available through the specific experiment and are further limited to low-dimensional signals. As such, they are ill-suited for the active learning of the structure-property relationships.

Here, we implement the deep kernel learning (DKL) based experimental workflow for active discovery of structure-property relationships in ferroelectric materials. This approach combines the power of machine learning to establish the correlative relationships between multidimensional data sets, and human based physical reasoning to establish targets for exploration based on observations and their uncertainties. Here, the relationship between the local domain structure and hysteresis loop is explored and future measurement locations are selected based on learnt relationships between local domain structure and polarization switching behavior.

To illustrate the principle of the DKL applications in experiment, we first implement DKL using a pre-acquired high density band excitation piezoresponse spectroscopy (BEPS) imaging data set, which hence provides the "known" ground truth image. Here, as a model system we have chosen a $PbTiO_3$ (PTO) thin film grown on (001) $KTaO_3$ substrates with a $SrRuO_3$ conducting buffer layer by metalorganic chemical vapor deposition (MOCVD) method, as reported by H. Morioka et al.[40] Shown in Figure 1a is the topography image of the PTO sample illustrating clear ferroelastic domain wall pattern. The clearly visible corrugations on the sample surface are associated with the ferroelastic domain walls between the domains with different polarization orientations. Here, the lattice mismatch across the single domain walls gives rise to strain and hence deformations. The superposition of deformations form multiple domain walls give rise to ripple-like structure.

Shown in Figure 1b-c are the corresponding band excitation (BE) PFM amplitude and phase images, which indicate the existence of both in-plane a domains and out-of-plane c domains. In the in-plane *a* domain, the polarization vector is parallel to the surface and hence associated electromechanical response amplitude is close to zero. At the same time, in the out of plane *c*-domains the response amplitude is high. Finally, the phase image indicates whether polarization vector is parallel or antiparallel to surface normal.

To explore polarization dynamics in this system, we analyze the high density polarization loop measurements.[41] In these measurements, the dc component of the tip bias is following the triangular waveform, inducing domain nucleation and growth below the tip. The resultant changes of the PFM signal are recorded as the local hysteresis loop. The shape of the hysteresis loop this reflects the mechanism of local polarization switching as affected by the ferroelastic walls, structural defects, etc. The resultant 3D data array can be analyzed to extract the descriptors of the hysteresis loop such as area under the loop, coercive and nucleation biases, etc. that can further be plotted as 2D maps. A high grid density (100x100) polarization image is show in Figure 1d, where the similar domain structure is visible. Shown in Figure 1e is a ferroelectric polarization hysteresis loop from the orange point marked on Figure 1d.

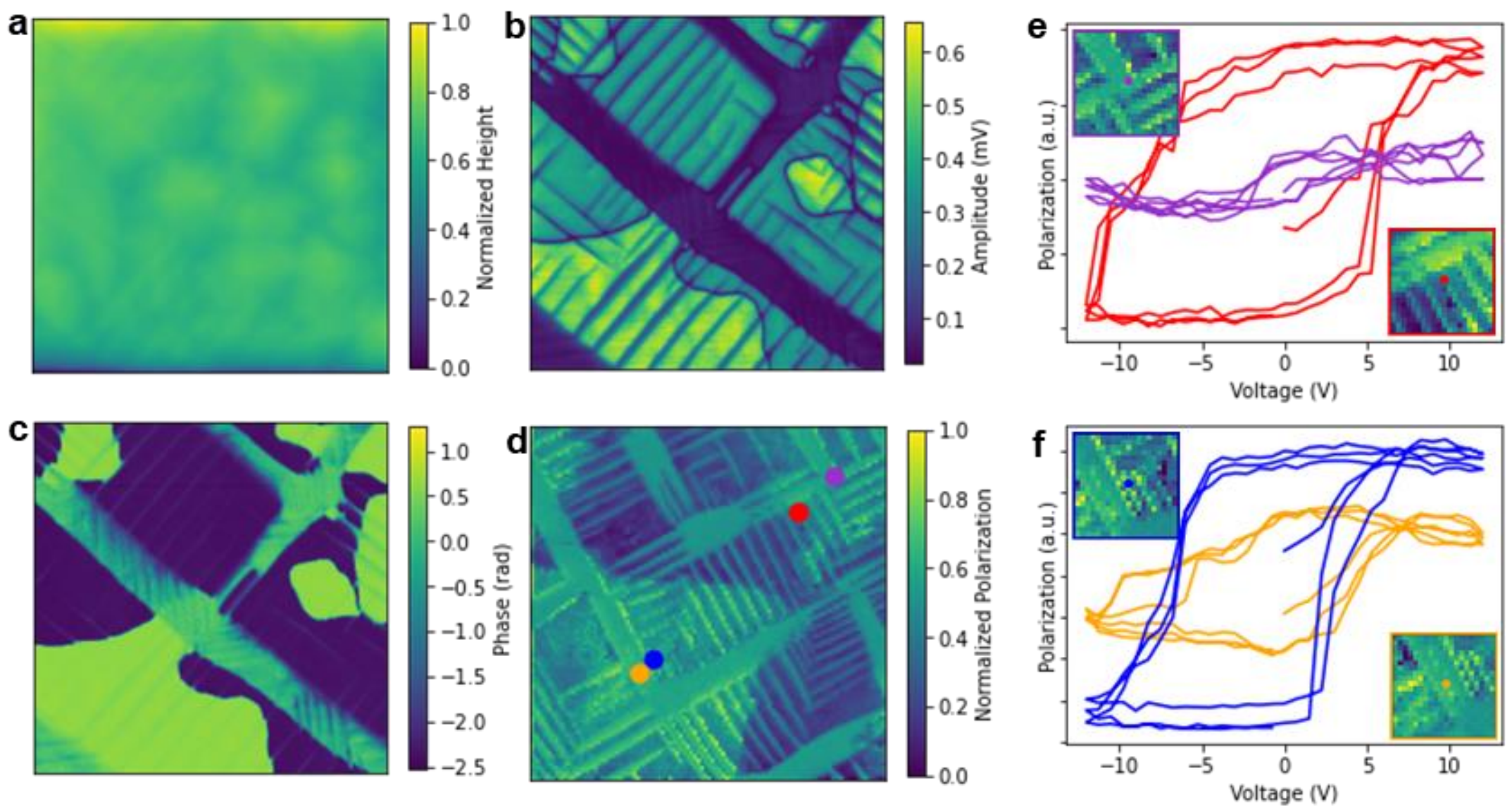

**Figure 1.** PFM results from PTO film. **(a)**, topography and corresponding band excitation **(b)** amplitude and **(c)** phase maps show c/c and a/c domains. **(d)**, BEPS polarization map when $V_{dc}$ = 0 V. **(e)-(f)**, BEPS polarization hysteresis loops from the locations labeled on (d); inserts in (e) and (f) show the domain structures around the locations where hysteresis loops are from, the correspondence between insert images and hysteresis loops is indicated by colors; these illustrate that different elements of hysteresis loops are correlated with different elements of domain structures. Note that ground truth BEPS image contains 10,000 hysteresis loops measured over uniform spatial grid.

As acquired, the combination of the PFM domain structure image and hysteresis loop mapping allows to reconstruct structure property relationships, defined here as a correlative link between the local ferroelastic domain structure and hysteresis loop shape. Previously, we have demonstrated the use of machine learning methods, specifically an encoder-decoder type of neural networks, to build such relationships for ferroelectric[42,43] and plasmonic[44] structures. These correlative relationships allow answering questions such as, what responses are possible in a given system, what structures are necessary to maximize certain aspects of the response, etc. However, as with any correlative method, these answers are valid only for the in-distribution data (meaning for the same material under the same microscope settings), and do not allow answering counterfactual and interventional questions.[45-47] Some of these can be answered via transition from correlative to generative physical models,[48-50] but these raise further questions of theory-experiment matching.

Furthermore, these analyses are limited to the case when the full data set is available *a priori*, i.e., they allow analyzing the data after the experiment. However, it is not guaranteed that the pre-acquired data sets sampled most interesting locations, and the number of possible measurements are limited by acquisition time and probe stability. Contrarily, in an ***active*** machine learning setting, only the full topographic and PFM images are available, and information contained in these images is used to select the locations for spectroscopic measurements. Based on the examination of spectroscopic data, further locations are identified. For example, the operator can learn that the in-plane $a$-domain regions do not have measurable hysteresis loop, that all $a$-$c$ domain structures have similar switching behaviors, and that irregular domain edges or junctions may possess interesting dynamics. The subsequent selection of the target locations can then be based on these observations and curiosity (exploration) or perceived usefulness (exploitation). Importantly, this approach is the basis of DKL.

The DKL is based on the Gaussian Process (GP) regression, and can be represented as a combination of GP with deep neural networks. In general, GP generally refers to an indexed collection of random variables, any finite subcollection of which have a joint multivariate Gaussian distribution.[51] A GP is completely determined by its mean and covariance functions, with the latter determining the functional form and strength of interaction between the points in the input space.

A common application of GP is in a regression setting where it can be used for reconstructing data from sparse observations with quantified uncertainty.[52] Note that the uncertainty per se is important when gaining quantitative insights into physical behaviors, and locations with high uncertainty can indicate the presence of new physical mechanisms.[53] Specifically, given the dataset $D = [x^i, y^i]_{i=1,\dots,N}$, where $x$ and $y$ represent inputs/features and outputs/targets, respectively, the GP probabilistic regression model with a standard squared exponential kernel of the form $k(\mathbf{x}, \mathbf{x}') = \sigma^2 \exp(\frac{1}{2}(\mathbf{x} - \mathbf{x}')^2/l^2)$ is defined as

$$y \sim MultivariateNormal\big(0, K(\mathbf{x}, \mathbf{x}', \sigma, l)\big) \tag{1a}$$

$$\sigma \sim LogNormal(0, s_1^{\text{const}}) \tag{1b}$$

$$l \sim LogNormal(0, s_2^{\text{const}}) \tag{1c}$$

where $K$ denotes a function that computes a kernel matrix such that $K_{ij} = k(x_i, x_j)$ for the sampled kernel hyperparameters. The GP model can be trained either using a Markov Chain Monte Carlo algorithm on the model to get posterior samples for the GP parameters or via a variational inference. It is commonly assumed that there is an observation noise such that $\mathbf{y}_{\text{noisy}} = \mathbf{y} + \boldsymbol{\varepsilon}$ where $\boldsymbol{\varepsilon}$ is a normally distributed noise with zero mean and variance $s^2_{\text{noise}}$. Practically, this noise gets absorbed into a computation of the covariance function.

Once the GP model parameters are learned, it can be used to make predictions on new "test" points. This is done by sampling from the multivariate normal posterior over the model outputs at the provided test points $x_*$:

$$f_* \sim MultivariateNormal\left(\mu^{\text{post}}_{\boldsymbol{\theta}}, \Sigma^{\text{post}}_{\boldsymbol{\theta}}\right) \tag{2a}$$

where $\boldsymbol{\theta} = [\sigma, l]$ and

$$\mu^{\text{post}}_{\boldsymbol{\theta}} = K(x_*, x|\boldsymbol{\theta})K(x, x|\boldsymbol{\theta})^{-1}\mathbf{y} \tag{2b}$$

$$\Sigma^{\text{post}}_{\boldsymbol{\theta}} = K(x_*, x_*|\boldsymbol{\theta}) - K(x_*, x|\boldsymbol{\theta})K(x, x|\boldsymbol{\theta})^{-1}K(x, x_*|\boldsymbol{\theta}) \tag{2c}$$

The GP predictive mean ($\bar{f}_*$) and variance ($\mathbb{V}[f_*]$) can be used to select the next measurement point(s) via a so-called acquisition function,[54] $acq(\bar{f}_*, \mathbb{V}[f_*])$, so that $x_{\text{next}} = \text{argmax}(acq)$. The acquisition function reflects the measure of interest to specific region based on expected function value and uncertainty, balancing exploration and exploitation. Implementation-wise, one generally starts with a few sparse observations and trains a GP model. Then, a prediction on the "test" points – which usually represent all the unmeasured points in a selected parameter space – is made and used to derive an acquisition function for sampling the next query point. This approach is referred to as Bayesian optimization (BO) or 'active learning'. For structural imaging in microscopy, the parameter space corresponds to a 2D grid over a chosen scan area; for spectroscopic measurement, a third dimension corresponding to the energy axis can be added.

A significant limitation of the standard GP-based active learning is that it does not scale well with dimensionality of the parameter space. As a result, for many common hyperspectral measurements in 3D-5D space, the GP training and inference may take so long (even on modern Graphics Processing Units) that it is faster to perform the measurement by simply sampling all the points (i.e., the standard way of doing measurements). Another limitation is that the standard GP does not, strictly speaking, learn representations of data which precludes from using prior knowledge from different experimental modalities to assist in the experiment (something that a (good) experimentalist does all the time). In the context of the ferroelectric domain studies detailed here, the simple GP/BO does not use the information on the preexisting domain structure to build the relationship between these and switching behaviors.

To address these issues, here we have adapted a deep kernel learning[55] approach where a neural network is used to convert high-dimensional input data into a set of low-dimensional descriptors on which a standard ('base') GP kernel operates (Figure 2a). Formally, we define our 'deep kernel' as

$$k_{DKL}(x, x'|\boldsymbol{w}, \boldsymbol{\theta}) = k_{base}(g(x|\boldsymbol{w}), g(x'|\boldsymbol{w})|\boldsymbol{\theta}) \quad (3)$$

where $\boldsymbol{w}$ are the weights of the neural network. Hence, the deep kernel operates in the latent (embedding) space learned by a neural network from the (potentially high-dimensional) data and can be referred to as the data-informed kernel. The parameters of neural network and GP base kernel are learned simultaneously by maximizing the model evidence via a stochastic variational inference.[56] The trained DKL GP model is then used for obtaining predictive mean and variances following the Eq. (2), that is, using the same procedure as for the standard GP. Then, the acquisition function for the expected improvement[54] is used to predict the next measurement point. The DKL can operate both on scalar (single output) and vector (multiple outputs) targets. In the latter case, different function outputs (such as response function values at different energies) can be independent or correlated. Implementation-wise, the correlation between different function outputs is achieved by forcing them to share the same latent space, i.e. having a single neural network (feature extractor) connected to multiple GPs whose number is equal to the number of function outputs. Alternatively, one may assume independence between outputs, which would require training an independent neural network for each GP (i.e., for each output). For a single-objective active learning, the vector-valued prediction of the DKL model must be scalarized in order to select the next measurement point.

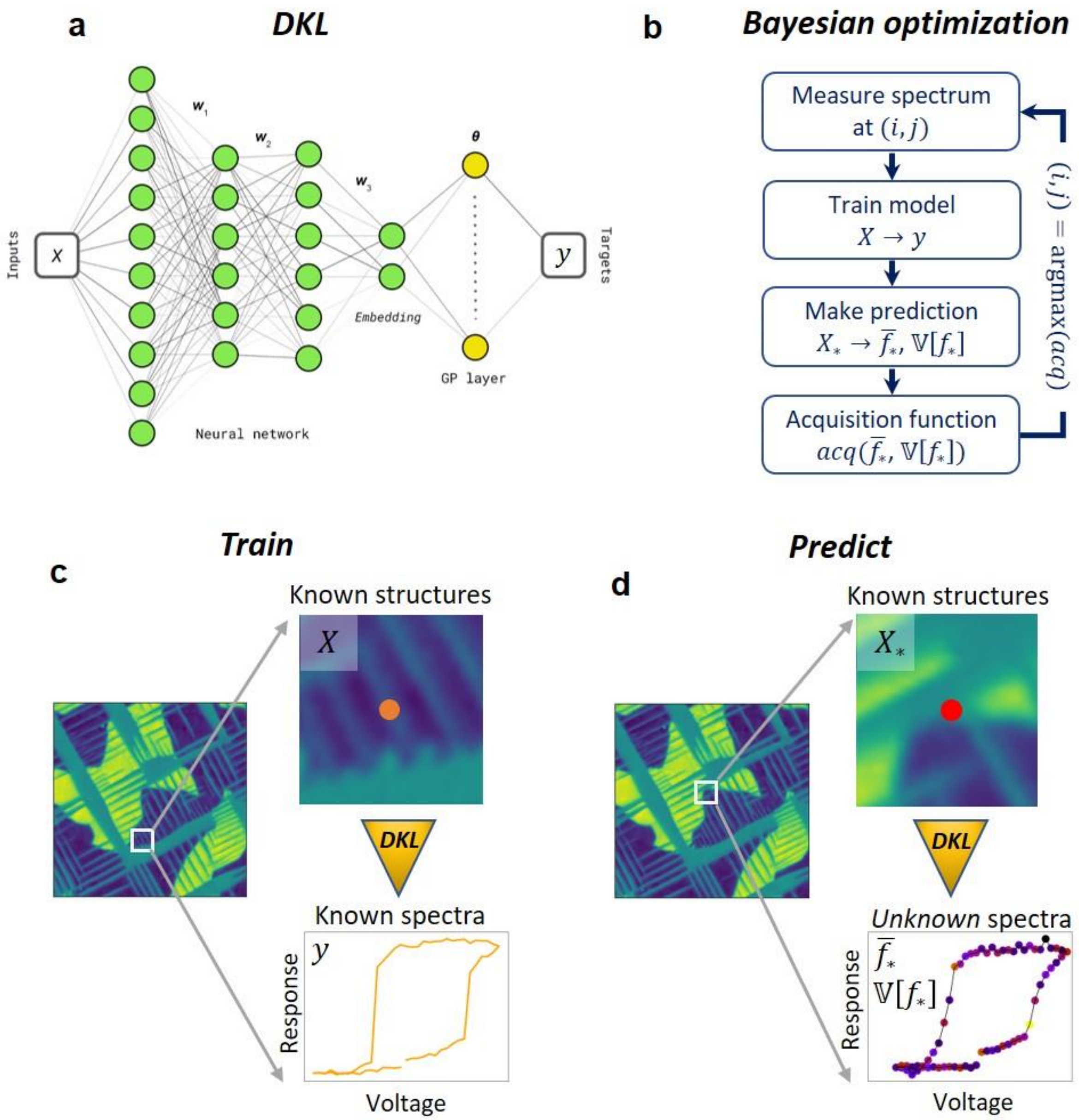

**Figure *2*.** Schematic illustration of active learning with deep kernel learning (DKL). Here, the algorithm has access to the full topographic data and uses Bayesian optimization to learn the relationship between the domain structure within the patch and hysteresis loop. (**a)** The inner structure of the DKL model. A feedforward neural network parametrized by weights $\boldsymbol{w}$ projects the potentially high-dimensional input data $X$ into the low-dimensional latent space, $g(x|\boldsymbol{w})$ in which a standard GP kernel $k(z, z'|\boldsymbol{\theta})$ operates (where $z$ are the embedded/latent features). (**b)** The Bayesian optimization loop. (**c**) At training step, the DKL-based Gaussian process (GP) regression model (for brevity, referred to as DKL model) is trained using a small number of observations where inputs are topographic image patches and outputs are corresponding spectra or scalar values of a specific property derived from the spectra. (**d)** At prediction step, a trained DKL model is used to predict spectra at *every* coordinate in the topographic image for which there is no measured

spectra. Importantly, this method outputs both expected mean value, $\bar{f}_*$, of the property of interest and the associated uncertainty, $\mathbb{V}[f_*]$, which are used to derive an acquisition function $acq(\bar{f}_*, \mathbb{V}[f_*])$ for selecting the next measurement point (see **b**). Note that if the output is a vector-valued function (such as spectra) it needs to be scalarized before passing to the acquisition function such that the exploration process is controlled by a single descriptor that is a function of the predicted functionality and its uncertainty. In this manner, the human operator defines what physical functionality is targeted during the experiment. Note that exploration can be based both on physical and on information-theoretical criteria, i.e. targeting variability of observed behaviors (curiosity learning).

To illustrate DKL approach, we first implement it on the pre-acquired data set. Here, we use the random sampling with the 15% of measured points for training the DKL model that subsequently makes predictions on the full dataset. Shown in Figure 3 are results of DKL analyses on the pre-acquired BEPS data (as shown in Figure 1d-e), where we show the ground truth, embedded variables, DKL prediction, and the associated uncertainty. Both analyses on hysteresis loop area and hysteresis loop width indicate good reconstructions of loop area and loop width maps (Figure 3d and 3i) when comparing with the ground truth maps (Figure 3a and 3f). The embedded variables highlight the tiny domains (Figure 3b and 3g) and the large domains (Figure 3c and 3h), respectively. Shown in Figure 3k-m are several examples demonstrating the reconstruction of hysteresis loops comparing with the ground truth hysteresis loops.

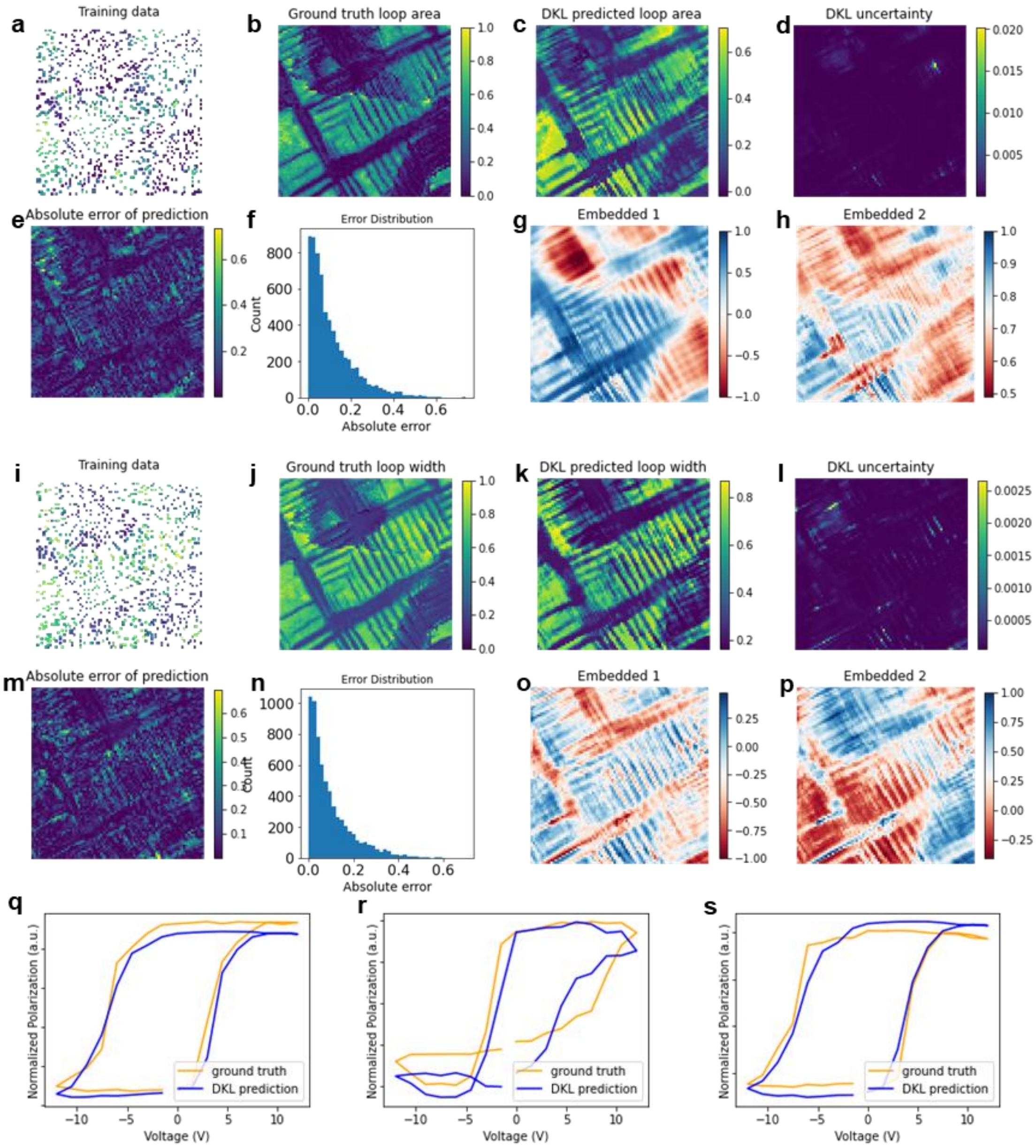

**Figure 3.** Deep kernel learning on the pre-acquired data set for 15% of measured locations. **(a-h)** DKL analysis of loop area: **(a)** 15 % randomly sampled loop area data for training session, **(b)** ground truth loop area map, **(c)** DKL predicted loop area, **(d)** DKL uncertainty map, **(e, f)** absolute error map of DKL prediction and the histogram distribution of errors, **(g, h)** the embedded latent maps of the trained DKL (the embedding dimensions were set to 2). **(i-p)** DKL analysis of loop width: **(i)** 15 % randomly sampled loop width data for training session, **(j)** ground truth loop width map, **(k)** DKL predicted loop width, **(l)** DKL uncertainty map, **(m-n)** absolute error map of DKL prediction and the histogram distribution of error, **(o-p)** the embedded latent maps of the trained DKL. **(i, j)** DKL predicted loop width and DKL uncertainty maps. **(q-s)** examples of DKL prediction on hysteresis loops, showing DKL predicted loops and ground truth loops.

We further explore the effect of the number of sampling points for random sampling. Here, we show the DKL reconstruction of coercive field, hysteresis loop area, and loop width based on 1%, 3%, 5 %, and 10% of random sample data in Figure 4a-c. Note that in this approach DKL aims to reconstruct the relationship between the local domain structure and the hysteresis loop properties based on fully known collection of image patches, and patch-spectrum pairs available for only (very small) fraction of the data. With this relationship established, it aims to reconstruct the spectra properties and their uncertainties for the full data set. Figure 4d-e show the quality of reconstruction as the mean squared error (MSE) between the reconstruction and the ground truth as a function of the number of sampled data points. Note that in Figure 4a, the coercive field is the average of positive and negative coercive field; in Figure 4b and 4c, the loop area and width are from off field hysteresis loops. More results about separated positive and negative coercive fields, and on field loop area are shown in Supplementary Information Figure S1. More details about how the random sampling of data affects the DKL reconstruction are shows in Supplementary Information as videos Supplementary Videos S1-S5.

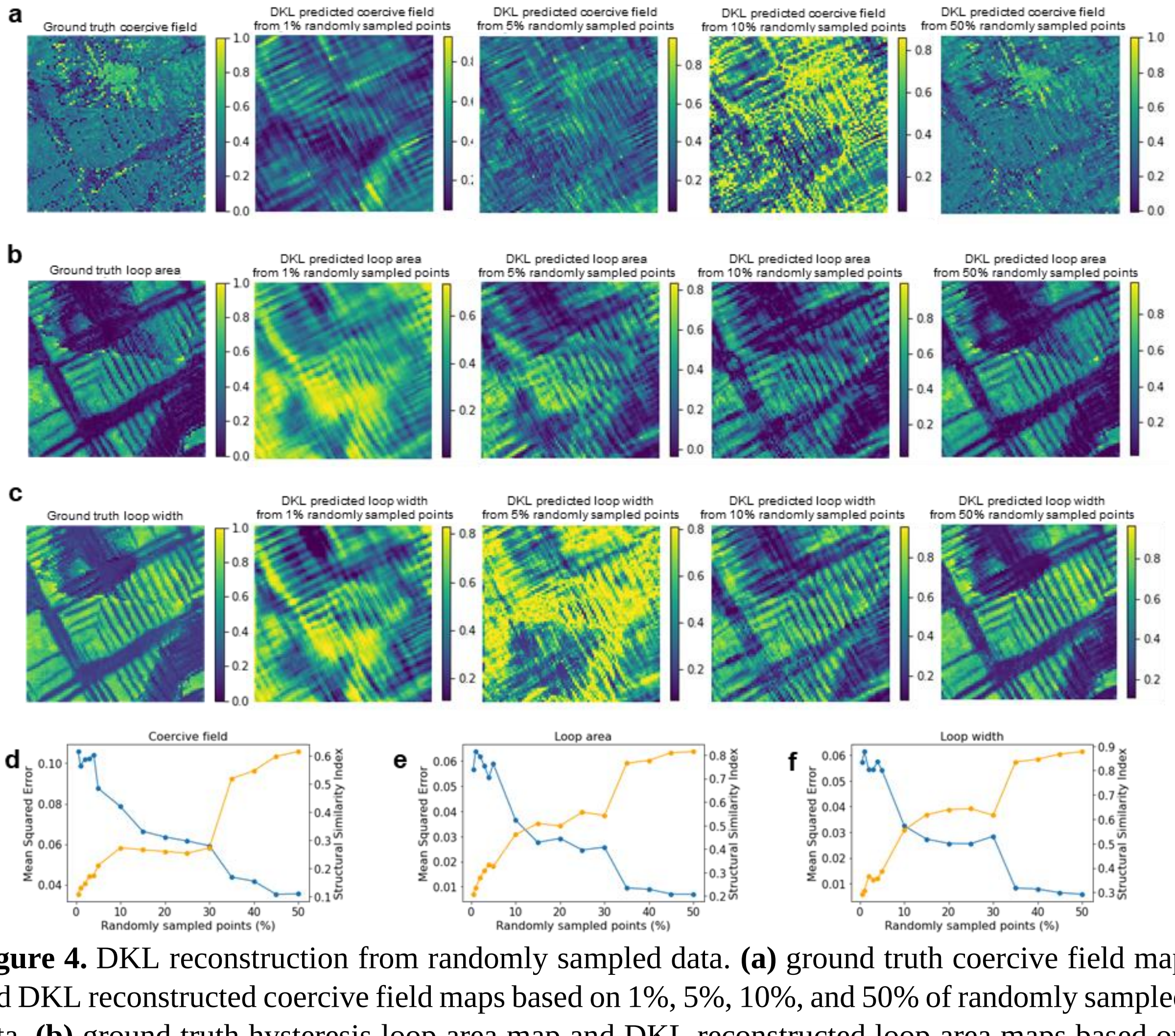

**Figure 4.** DKL reconstruction from randomly sampled data. **(a)** ground truth coercive field map and DKL reconstructed coercive field maps based on 1%, 5%, 10%, and 50% of randomly sampled data. **(b)** ground truth hysteresis loop area map and DKL reconstructed loop area maps based on 1%, 5%, 10 %, and 50% of randomly sampled data. **(c)** ground truth hysteresis loop width map

and DKL reconstructed loop width maps based on 1%, 5%, 10 %, and 50% of randomly sampled data. **(d-f)** DKL reconstruction error and structure similarity between ground truth and reconstruction of coercive field map, loop area map, and loop width map, respectively. Note that reconstructed images preserve features at all length scales, which very difficult to achieve with classical BO.[39]

We further illustrate the transition from the reconstruction based on predefined (e.g. random or low-density grid) sampling towards science-driven discovery, and experimental implementation of this approach. Here, the critical new element is the definition of the scalar descriptor that reflects the physical behaviors we are interested in and use to guide the exploration process. For hysteresis loops a shown in Figure 5a, these can be parameters such as loop area or width, coercive field, and nucleation bias, or more complex descriptors such as quality of functional fit by predefined function, fit parameters, etc. Alternatively, the exploration can be based on information-theoretical criteria such as growth of entropy of the data set, i.e. curiosity learning. The key element here is that the acquisition function summarizing the degree of interest to specific behavior allows human-level decision making searching for specific physical signatures, and allows to incorporate associated uncertainties.

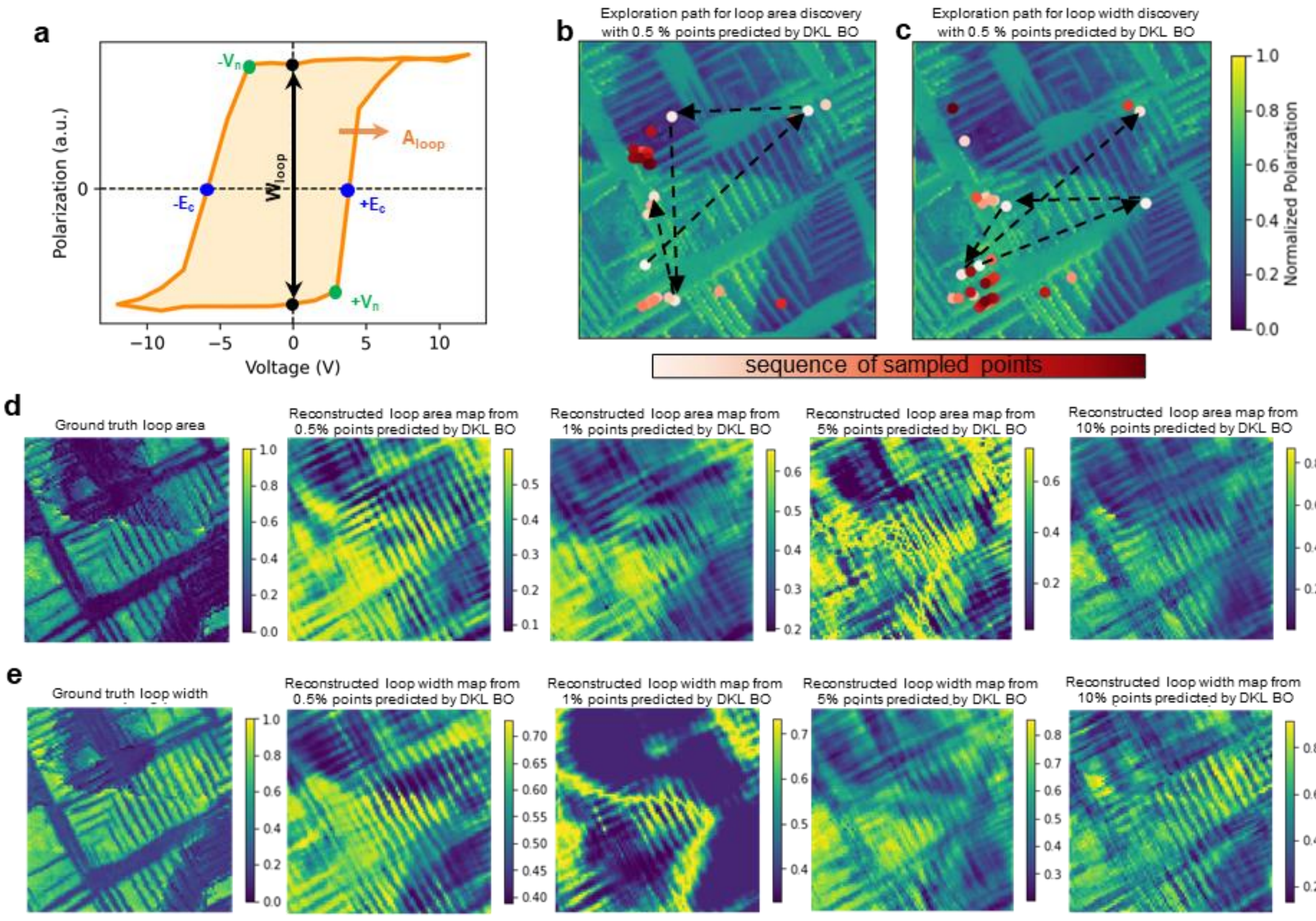

**Figure 5.** Physics-based descriptors for DKL BO. **(a)** example of hysteresis loop and definition of possible physical descriptors, $A_{loop}$: loop area; $W_{loop}$: loop width; $-E_c$: negative coercive field; $+E_c$:

positive coercive field; $-V_n$: negative nucleation bias; $+V_n$: positive nucleation bias. **(b)** exploration path for loop area discovery. **(c)** exploration path for loop width discovery. **(d)** ground truth loop area map and DKL BO reconstructed loop area maps based on loop area discovery for 0.5 %, 1 %, 5 %, and 10 % sampling points. (e) ground truth loop width map and DKL BO reconstructed loop area maps based on loop width discovery for 0.5 %, 1 %, 5 %, and 10 % points predicted by DKL BO. The associated videos and the behavior of uncertainty are available in the supplementary materials.

Shown in Figure 5b-c are the DKL BO navigation sequence with the acquisition function based on hysteresis loop area and loop width, respectively. Shown in Figure 4d-e are the DKL reconstruction of loop area and width with 0.5 %, 1 %, 5 %, and 10 % of the points where the DKL BO suggested to perform measurements. The reconstruction quality can be evaluated by comparison with the random sampling points in Figure 4. We show the videos of acquisition function image with labels of discovered points in Supplementary Information for both loop area and loop width cases. The reconstruction videos with the DKL-BO sampled points, DKL prediction, and DKL uncertainty are also shown in Supplementary Information.

Finally, we deploy the DKL discovery workflow on the operational microscope. Here, we combined the DKL discovery workflow in Jupyter notebook with an in-house LabView-based script for National Instruments hardware (LabView-NI) to control the tip position for BEPS waveform generation and data acquisition. First, we performed a BEPFM measurement to acquire the domain structure image, which will be used to generated domain structure image patches for DKL. At the beginning of BEPS hysteresis loop measurement, the sampling point is initialized by the LabView-NI at a random location to obtain hysteresis loop. Then, the Jupyter notebook analyzes the hysteresis loop and the corresponding pre-acquired domain structure image patch to train the DKL model. The DKL was trained for 200 iterations after which a prediction on all the image patches was made and the pre-selected acquisition function was used to derive the next location for hysteresis measurement to LabView-NI. Then the process is repeated. A schematic of this specific workflow used is shown in Figure 6a. Figure 6b-c shows BEPFM amplitude and phase images with 256*256 grid size which will be used to generate image patches for DKL. It can be seen that the PTO film contains both in-plane *a* domains and out-of-plane *c* domains as demonstrated previously. The size of the generated image patches for DKL-BO is 20*20 grid and the BEPS measurement was performed at 237*237 grid size. The DKL-BO process was continued for 200 steps, i.e., until 200 pixels worth of data were captured.

We used the image patches generated from the BEPFM results in Figure 6b and 6c for guide the discovery workflow using on-field hysteresis loop area and off-field hysteresis loop area, respectively. The discovered points are labeled in Figure 6b-c. The detailed discovery processes are shown in Supplementary Information as videos of acquisition function images with labeled exploration point. Interestingly, the DKL BO sampled points for on-field hysteresis loop area are concentrated around $c^-/c^+$ ferroelectric domain walls (Figure 6b), while the DKL-BO sampled points for off-field hysteresis loop area are concentrated around a/c ferroelastic domain walls, demonstrating the potential of this approach to discover different behaviors based on predefined exploration targets. This is an indicative of the different properties included in the on-field and off-field hysteresis loops. With the obtained 200 hysteresis loops and the BEPFM domain structure images, we are able to make predictions on the hysteresis loop area maps. Shown in Figure 6d-e are the DKL prediction of on-field and off-field hysteresis loop area maps, respectively, along with the DKL uncertainty. The domain structures are visible in the predicted loop area maps, indicating the hysteresis loop is associated with the domain structure.

Here, these observations can be readily rationalized (but not predicted!) based on the known physics of ferroelectric domain walls. Here, the larger polarization mobility in the vicinity of the 180 walls results in more significant hysteresis loop opening in the on-field measurements. At the same time, the off-field measurements detect only the slowly relaxing (on the measurement time scale) components, indicative of the stronger pinning at the ferroelastic walls. These behaviors generally agree with some of the prior observations of similar systems.[57,58]

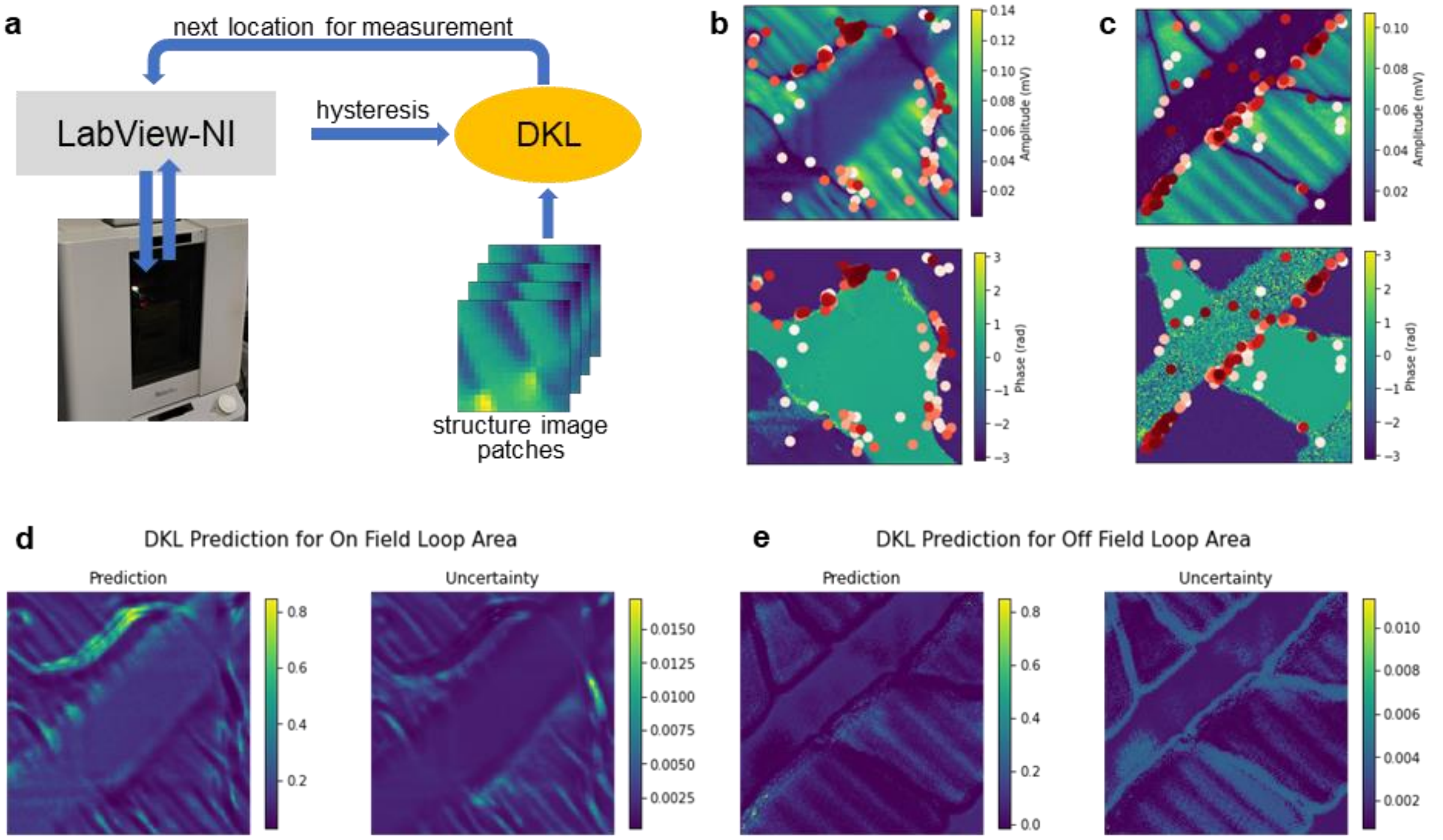

**Figure 6.** DKL-BO based automated PFM experiment. **(a).** schematic of the DKL-BO PFM workflow. **(b)-(c)** BEPFM amplitude and phase images used for generation of domain structure image patches for DKL-BO BEPS measurements; the BEPFM results in (b) and (c) are used for DKL-BO discovery of on-field loop area and off-field area, respectively; the discovered points are labeled in the image in (b) and (c). **(d)-(e),** DKL prediction of on-field and off-field loop area maps based on the obtained 200 hysteresis loops.

To summarize, we have developed the DKL approach that allows the physical discovery in automated experiment. Compared to the classical Bayesian Optimization based strategies that use a single (or small number) of scalar descriptors to guide the navigation process and do not incorporate the prior knowledge, this approach uses the data contained in structural images to identify the locations of the spectroscopic measurements, and identifies new locations and builds the structure property relationships simultaneously. This discovery process is guided by the acquisition function that is constructed from predicted behavior and its uncertainty, and reflects the target of the experiment. This target can be optimization of specific property, similarity to a given model, or novelty discovery. In this manner, we combine the power of correlative machine learning methods to establish relationships between multidimensional data set and derive corresponding uncertainties, and human physics-based decision making encoded in the choice of the acquisition function.

Here, we implemented this approach for PFM measurement to investigate the relationship between polarization hysteresis and ferroelectric/ferroelastic domain structures. The obtained results show different exploration path and sampled points when the DKL is guided by on-field and off-field hysteresis loops, indicating structure-hysteresis relationship varies under different circumstances, i.e. on-field or off-field. We also note that in principle the DCNN part of the DKL can be pre-trained on previous experimental data from the same or similar systems, somewhat equivalent to the transfer learning approach. However, this will necessitate the stringent analysis of the out of distribution drift effects (e.g. due to different microscope settings).

Similarly, this workflow can be readily extended to other SPM modalities, including current-voltage curve measurements or relaxation measurement. We expect that most significant benefit will be achieved for measurements with the readily identifiable connection to materials physics (e.g. signature of Majorana fermions in STM), large acquisition times, and especially destructive measurements such as nanoindentation and irreversible electrochemical measurements. Beyond SPM, similar approaches can be used to techniques such electron microscopy, chemical and mas-spectroscopic imaging, and nanoindentation and micromechanical testing. Finally, the DKL approach can be implemented over more complex parameter spaces, e.g. for material discovery in combinatorial spread libraries or molecular systems.

## Methods

*Data analysis*

The detailed methodologies of DKL analysis on pre-acquired data are established in Jupyter notebooks and are available from https://git.io/JRspC. A standard MLP with three hidden layers containing 1000, 500, and 50 “neurons” was used as a ‘feature extractor’ in the DKL model.

*PTO sample*

The PTO film was grown by chemical vapor deposition on a $SrRuO_3$ bottom electrode on a $KTaO_3$ substrate.

*BEPFM and BEPS measurements*

The PFM was performed using an Oxford Instrument Asylum Research Cypher microscope with Budget Sensor Multi75E-G Cr/Pt coated AFM probes (~3 N/m). Band excitation data are acquired with a National Instruments DAQ card and chassis operated with a LabView framework.

*DKL-PFM implementation*

The DKL deployment notebook for BEPS measurement is available from https://git.io/JRspC, which can be adapted for other modalities.

## Conflict of Interest

The authors declare no conflict of interest.

## Authors Contribution

S.V.K. conceived the project and M.Z. realized the DKL-BO workflow. Y.L. performed detailed analyses with basic workflow from M.Z. Y.L. deployed the DKL to PFM measurement and obtained results. R.K.V. and K.K. helped with the deployment. H.F. provided the PTO sample. All authors contributed to discussions and the final manuscript.

## Acknowledgements

This effort (ML) was supported as part of the center for 3D Ferroelectric Microelectronics (3DFeM), an Energy Frontier Research Center funded by the U.S. Department of Energy (DOE), Office of Science, Basic Energy Sciences under Award Number DE-SC0021118 (Y.L., K.K., S.V.K.), and the Oak Ridge National Laboratory’s Center for Nanophase Materials Sciences (CNMS), a U.S. Department of Energy, Office of Science User Facility (M.Z., R.K.V).

## Data Availability Statement

The data that support the findings of this study are available at https://git.io/JRspC.